\newcommand{\be}{\begin{equation}}
\newcommand{\ee}{\end{equation}}
\newcommand{\bea}{\begin{eqnarray}}
\newcommand{\eea}{\end{eqnarray}}

\documentclass[prd,epsf,nofootinbib]{revtex4}
\usepackage{indentfirst}
\usepackage{latexsym}
\usepackage{graphicx}

\begin{document}

\baselineskip 18pt

\title{Assessing the Effects of the Uncertainty in Reheating Energy Scale on Primordial Spectrum and CMB  }
\author{Bo Feng,${}^{1}$ Xiaobo Gong,${}^{1,2}$  Xiulian Wang${}^{1}$\\
{\small $^{1}$ Institute of High Energy Physics, Chinese Academy of Science, P.O. Box 918-4, Beijing 100039, P.R.
China}\\
{\small $^{2}$ Department of Physics, Peking University, Beijing
100871, P.R. China}}

\begin{abstract}
The details of reheating energy scale $\rho_{reh}$ is largely
uncertain today, independent of inflation models. This would
induce uncertainty in predicting primordial spectrum. Such
uncertainty could be very large, especially for spectra with large running $n_S$.
We find that for some inflation models with a large
$d\ln n_S(k)/d\ln k$, $\rho_{reh}$ could be highly restricted by current CMB observations.
\end{abstract}

\maketitle

The cosmic microwave background radiation(CMBR) anisotropy for the
$l$-th multipole $C_l$ by definition is related to
 the angular correlation function\cite{Ma}: \be
\langle\Delta(\vec{n}_1)\Delta(\vec{n}_2)\rangle \equiv
\frac{1}{4\pi} \sum
\limits_{l=0}^{\infty}(2l+1)C_lP_l[\cos(\vec{n}_1\cdot\vec{n}_2)],
\ee
 and for scalar
temperature modes \be C_l^{S}\equiv
\frac{2\pi}{l(l+1)}\tilde{C_l}= \frac{4\pi}{(2l+1)} \int
\frac{dk}{k}T_l^2(k)P_S(k),
 \ee
where $T_l(k)$ is the transfer function and $P_S(k)$ is the
primordial spectrum from inflation. The scalar spectrum index is
defined as $n_s(k)\equiv 1+d\ln P_S(k)/d\ln k$.

Usually one can specify the $k$ modes during inflation basing on
the e-folds number $N(k)$ before the end of inflation\cite{Lyth
reports}:
 \bea
N(k)=
 62-\ln\frac{k}{a_0H_0}-\ln\frac{10^{16}GeV}{V(k)^{1/4}}
+\ln\frac{V(k)^{1/4}}{V_{end}^{1/4}}-
\frac{1}{3}\ln\frac{V_{end}^{1/4}}{\rho_{reh}^{1/4}}, \eea
 where $V(k)$, $V_{end}$ denote  the inflaton potential at $k=aH$
 and at the end of inflation respectively, $\rho_{reh}$ is the energy density
  when reheating ends, resuming a standard big bang
 evolution. Commonly one takes  $a_0=1$, where
the subscript $'0'$ denotes the present value. For a given model
of inflation(which ends naturally), $P_S(a/a_{end})$ or $P_S(N)$
can be exactly formulated, however $N(k)$ cannot be specified
accurately in the lack of details of reheating. Consequently , we
are unable to tell precisely when a mode $ k $ leaves out horizon
, and there exists an uncertainty in predicting the primordial
spectrum and its effects on $\tilde{C_l}$ . In this paper we study
in detail two effects related to the uncertainty of $N(k)$ : the
normalization of inflaton's potential and the effects on the
spectrum of CMBR.

Commonly one assumes $n_s = constant$ without using specific
inflation models when fitting cosmological parameters to
observations. An analysis by Wang $et$ $al.$ \cite{Xiaom Wang}
gave that current observations indicated
$n_s=0.91^{+0.15}_{-0.07}$. However,for a general inflation model,
it does not give exactly $n_s = constant$. Sometimes significant
scale-dependence of $n_s(k)$ is possible and allowed by current
data\cite{L.Covi,Wangxiulian,Covi02}. It is often important to
relate primordial spectra from specific inflation models with
CMBR observations for constraining model parameters, making
distinctive predictions and excluding some certain inflation
models.

When applying $k$ to $P_S$, a typical method is to specify a mode
around the COBE  pivot scale $k_ {COBE}\approx
7.0a_0H_0$\cite{Bunn Liddle and White}. The three undecided energy
scales $V(k_{COBE})$, $V_{end}^{1/4}$ and $\rho_{reh}^{1/4}$ give
a typical range of
 $N(k_ {COBE})= 40 \sim 60$\cite{Lyth reports}, it is even possible that $N(k_{COBE})=0 \sim 60$
\cite{L.Covi} if there exists thermal inflation\cite{Thermal
inflation}. The fit to CMB data can give an
 estimation of $V(k_{COBE})$\cite{Bunn Liddle and White} for specific inflation models\cite{L.Covi}. However
 the reheating energy scale cannot be restricted
 stringently  by observations today. Phenomenologically $\rho_{reh}^{1/4}$
 can be in the range of $1 MeV \sim 10^{16} GeV$, which based on Eq.3, gives rise to an uncertainty of 15 in $N_{COBE}$ .

If the slow rolling(SR) parameter $\epsilon$ is small enough, the
effect of changing $N(k_{COBE})$ is exactly like a horizontal
translation in the $P_S(k)-k$ picture. When the primordial
spectrum is scale invariant, the translation changes nothing. As
long as $P_S(k)$ is tilted, choosing $a_1H_1$ or $a_2H_2$ as
$k_{COBE}$ would globally change the value of $P_S(k)$. When
$n_s(k)$ is scale independent, different choice of $k_{COBE}$ will
change $P_S(k)$ to $c P_S(k)$, where $c$ is a constant, acting
also like a vertical shift of $P_S(k)$ in the $P_S(k)-k$ picture.
In Fig.1 we show the effect of choosing different $N_{COBE}$.
Assuming the Hubble constant $H$ remains the same on the shown
scale, although $P_S(N)$ is exactly known, $N_{COBE}=50$ or 60
will give different value for $P_S(k)$. Assuming constant $n_s$
and no gravitational waves, Bunn and White's fitting to COBE
observation gives\cite{Bunn and White}:
\begin{equation}
\left|\delta_{H}(k_{COBE})\right|=1.94 \times 10^{-5}\times
\Omega_0^{-0.785-0.05 \ln
\Omega_0}\exp[-0.95(n_s-1)-0.169(n_s-1)^2],
\end{equation}
  with an uncertainty less than $10 \%$ , where
\be
\delta_H(k)=\frac{2}{5}\frac{g(\Omega_0)}{\Omega_0}P_S^{1/2}(k)
\ee and \be
g(\Omega_0)=\frac{5}{2}\Omega_0(\frac{1}{70}+\frac{209\Omega_0}{140}-\frac{\Omega^2_0}{140}+\Omega_0^{4/7})^{-1},
\ee  with less than 5\% uncertainty for $g(\Omega_0)/ \Omega_0$.
$P_S(k_{COBE})$ has been stringently restricted for given $n_s$
and $\Omega_0$. On the other hand, taking single-field inflation
model $V=V_0 f(\phi/ \mu)$ as a example, one has \be P_S(k)
\propto \frac{V^3}{V_{\phi}^2} \propto V_0  . \ee For different
$N_{COBE}$, $ V_0 $ has to be chosen  differently to match the
COBE normalization.
 For $n_s=0.84$, $\Delta N=15$ the uncertainty on the amplitude of
 $V_0$ is
about $e^{15(1-0.84)}\approx 11$ .

In some sense the uncertainty of $V_0$ matters little, since
current inflation theories can not physically give exact value of
$V_0$. $V_0$ can even be used as a free parameter to be normalized
by CMB and LSS observations. For $n_s = constant$, the effects
caused by the uncertainty of $N_{COBE}$  will be fully cancelled
by normalization. However, in the case of $dn_s/d \ln k\not = 0$,
even if $V_0$ is best used, two normalized primordial spectra
$P_S(k)$ with different choosing of $N(k_{COBE})$ still cannot
fully overlap.

 Now let us study the effects on CMB. For simplicity we have fixed the
cosmological parameters as $h=.64,\Omega_{\Lambda}=0.66,\Omega_b
h^2=0.020$ and
 $\Omega_k=0$\cite{Xiaom Wang} in the following studies.

Firstly when the running of $n_S$ is very small, the uncertainty
of $\rho_{reh}$ cannot bring forth observable effects in CMB. In
Fig.2 we give an example \be
 P_S(k)=A(k/k_{\star})^{0.91-1-0.001\ln(k/k_{\star})}    ,
\ee where the constant $A$ is to be normalized. $C_{0l}$ denotes
$C_{l}$ when setting $N_{COBE}= N_{\star}$. For assessing the
difference, we have used the cosmic variance values, where
\cite{white93,ZS97,KAA97}
 \be
 \Delta \tilde{C_l} = \sqrt{\frac{2}{2l+1}} \tilde{C_l}     .
 \ee
The two specified lines stand for choosing
$N_{COBE}^{\prime}=N_{\star}+15$ and $N_{\star}-15$ respectively.
They are both within the cosmic variance of $C_{0l}$ for
$l\lesssim 2000$ and are hard to distinguish from $C_{0l}$. We've
also checked the effect of choosing different $n_{\star}$ in Eq.8,
and find the result remains the same for $0.84<n_{\star}<1.06$. In
this sense, for many inflation models in which $dn_s/dlnk $ are
very small, such as chaotic and natural inflation, the effect
caused by the uncertainty of reheating temperature is negligible,
provided $V_0$ can be normalized.

Secondly we rebuild a primordial spectrum with a constant running
$dn_S/d \ln k=-0.022$:
 \be
 P_S(k)=A(k/k_{\star})^{1.014-1-0.011\ln(k/k_{\star})}    .
 \ee  $n_S$ runs from 1.06 to 0.84 on the scale relevant to CMB
and LSS observations($k=3\times 10^{-4} \sim 6 $ $h Mpc^{-1}$)
\cite{TZ02} when choosing $k_{\star}$=$k_{COBE}$.
 $P_S(k)$ with constant index $n_S$=$0.91^{+0.15}_{-0.07}$ is allowed
 by current observations\cite{Xiaom Wang},
phenomenologically such a running should also satisfy current
observations.
 In Fig.3, $C_{0l}$ stands for $C_{l}$ when choosing
$k_{COBE}= k_{\star}$, and the two dashed lines are its cosmic
variance limits. One can see $C_{0l}$ is hardly distinguishable
from the constant index spectrum case in which $n_S=0.97$. As the
figure also shows, one e-fold of uncertainty in $N_{COBE}$ can not
make difference out of cosmic variance if $A$ can be best
normalized. And it is obvious that if $N_{COBE}$ varies larger
than 10, i.e. when $n_S$ is globally larger than 1.06 or less than
0.84, this spectrum will be excluded by current observations.

 Significant scale dependent primordial spectrum can explain the
 tentatively
observed feature at $k \sim 0.05 Mpc^{-1}$ \cite{LPS98,GSZ00}, and
in some degree solve the small scale problems of the $CDM$ model
on dark halo densities and dwarf satellites\cite{KL00}, but it
also can be easily excluded by current observations if $N(k)$
changes. The location of bump on $P_S$ may be shifted to different
$k$ by the uncertainty of $N_{COBE}$, 15 e-fold of uncertainty can
make the bump out of the range of CMB and LSS observation scale,
being fully unobservable, or to any other mode where $k\neq 0.05
Mpc^{-1}$, making the primordial spectrum disagree with the LSS
and CMB observations. In the BSI \cite{Starobinsky92} model which
has been used to solve the $ CDM$ puzzle on small scales, one
 requires a red-tilted $P_S(k)$ which is near scale invariant at
scale $k=3\times 10^{-4} \sim 6 $ $h Mpc^{-1}$ and damps severely
at larger $k$, however the location of such feature is still
sensitive to  $\rho_{reh}$.

Now let us study a specific inflation model considered in ref.
\cite{Wangxiulian}: \be V(\phi)=V_0(1+\cos \frac{\phi}{f}+\delta
\cos \frac{N\phi}{f}). \ee  For the calculations here we've set
$N=300$, $\delta= 5 \times 10^{-5}$ and $f=0.4 M_{Pl}$. Slow
rolling is well satisfied for $N> 50$, using the SR parameter
$\epsilon$: \be \epsilon = \frac{M_{Pl}^2}{16 \pi
 }(\frac{V_{\phi}}{V})^2           ,
 \ee
where $V_{\phi} \equiv \partial V / \partial \phi$. One can easily
get
 \be
 P_S(k)= \frac{8}{3 M_{Pl}^4} V(k)/\epsilon(k)          .
 \ee
With the background values of the cosmological parameters
mentioned above, COBE normalization predicts \be
V(k)^{\frac{1}{4}} \approx 6.4 \epsilon^{\frac{1}{4}}
\exp[-0.95(n_s-1)/2-0.169(n_s-1)^2/2] \times 10^{16} GeV
 \ee
 around $k_{COBE}$. For $ 0.84< n_S <1.06 $ the exponential factor
 is negligible. The normalization to full current CMB and LSS data differs little
with the COBE normalization. In Fig.4  the Hubble constant $H$
remains almost constant for $N> 40$ and it changes no more than
one percent on the full shown scale, rendering $dlnk \approx dN$.
$V(N)$ is also almost constant for $N> 40$ and
$V^{1/4}(N>40)/V^{1/4}_{end} \approx 2.0$, disregarding the exact
values of $V_0$. In the current CMB and LSS observation scale, the
minimum value of k is about $k_0 = H_0$, the maximum possible
value $N(k_0)$ is
 \be
N(k_0)\approx 62 - \ln(\frac{10^{16} GeV}{6.4\times 10^{14}GeV})
+\ln2 \approx 60.0
 \ee
where $\epsilon \sim 10^{-8}$. The corresponding spectrum index is
$n_S(k_0)\approx 0.96$. As one can see from Fig.4, the primordial
spectrum with a certain range where $N(k_0)>60.0$ should also fit
current CMB data well,
 but the details of this model has excluded the possibility. In the
figure, $N=52.6$ corresponds to $n_S =0.84$. It is also obvious
that for $n_S(k_0)\leq 0.84$, the model is excluded by current
observations, and $n_S(k_0)> 0.84 $ gives $\rho_{reh}^{1/4}
> 10^5 GeV$. To avoid too many
gravitinos, one requires $\rho_{reh}^{1/4} < 10^{10}
GeV$\cite{Sarkar}, i.e. $N(k_0)<56.5$ , $n_S(k_0)\leq 0.91 $.
 Phenomenologically when thermal inflation follows,
$N(k_0)$ can be much smaller, however, such spectra cannot
satisfy current CMB data, as shown in Fig.5. Only a small range
between the two dotted lines may be acceptable. In this range the
spectrum has a large running  $|dn_S/ d\ln k|\sim  0.01$, such an
$n_S$ cannot be excluded by current CMB data, as can be seen from
Fig.6 that it performs exactly like the constant $n_S=0.9$ case.
Meanwhile such a tilted spectrum exhibits the possibility
 of solving the CDM puzzle on small scales.

In summary, the uncertainties of $ \rho_{reh}$  has induced some
difficulty in formulating $P_S(k)$ from inflation model. This led
to additional degeneracy in the inflation parameter $V_0$ which
otherwise could be highly restricted for specific inflation
models\cite{TM TT}. For  models in which $dn_s/d\ln k$ is small
enough, the uncertainty is negligible. However, for models where
the spectrum is significant scale-dependent, the difference
induced by reheating can be very large as shown in this paper.
Furthermore, for some specific inflation models which have large
running $n_S$,
 current observations provide a constraint
on the reheating temperature. Before concluding we should point
that some aspects relevant to the uncertainty have been partly
mentioned in Refs\cite{Bunn Liddle and White,
Chung00,L.Covi,BCLH02}, but our paper is different from those
mentioned above since we mainly focus on the uncertainty of
reheating and on restriction of $\rho_{reh}$.

We thank  Dr. Mingzhe Li and Prof. Xinmin Zhang for discussions.
We also acknowledge the using of CMBfast code\cite{Seljak}. This
work is supported in part by National Natural Science Foundation
of China and by Ministry of Science and Technology of China.

\newcommand\AJ[3]{~Astron. J.{\bf ~#1}, #2~(#3)}
\newcommand\APJ[3]{~Astrophys. J.{\bf ~#1}, #2~ (#3)}
\newcommand\APJL[3]{~Astrophys. J. Lett. {\bf ~#1}, L#2~(#3)}
\newcommand\APP[3]{~Astropart. Phys. {\bf ~#1}, #2~(#3)}
\newcommand\CQG[3]{~Class. Quant. Grav.{\bf ~#1}, #2~(#3)}
\newcommand\JETPL[3]{~JETP. Lett.{\bf ~#1}, #2~(#3)}
\newcommand\MNRAS[3]{~Mon. Not. R. Astron. Soc.{\bf ~#1}, #2~(#3)}
\newcommand\MPLA[3]{~Mod. Phys. Lett. A{\bf ~#1}, #2~(#3)}
\newcommand\NAT[3]{~Nature{\bf ~#1}, #2~(#3)}
\newcommand\NPB[3]{~Nucl. Phys. B{\bf ~#1}, #2~(#3)}
\newcommand\PLB[3]{~Phys. Lett. B{\bf ~#1}, #2~(#3)}
\newcommand\PR[3]{~Phys. Rev.{\bf ~#1}, #2~(#3)}
\newcommand\PRL[3]{~Phys. Rev. Lett.{\bf ~#1}, #2~(#3)}
\newcommand\PRD[3]{~Phys. Rev. D{\bf ~#1}, #2~(#3)}
\newcommand\PROG[3]{~Prog. Theor. Phys.{\bf ~#1}, #2~(#3)}
\newcommand\PRPT[3]{~Phys.Rept.{\bf ~#1}, #2~(#3)}
\newcommand\RMP[3]{~Rev. Mod. Phys.{\bf ~#1}, #2~(#3)}
\newcommand\RPP[3]{~Rep. Prog. Phys.{\bf ~#1}, #2~(#3)}
\newcommand\SCI[3]{~Science{\bf ~#1}, #2~(#3)}
\newcommand\SAL[3]{~Sov. Astron. Lett{\bf ~#1}, #2~(#3)}

\newpage
\begin{figure}
\includegraphics[scale=0.4]{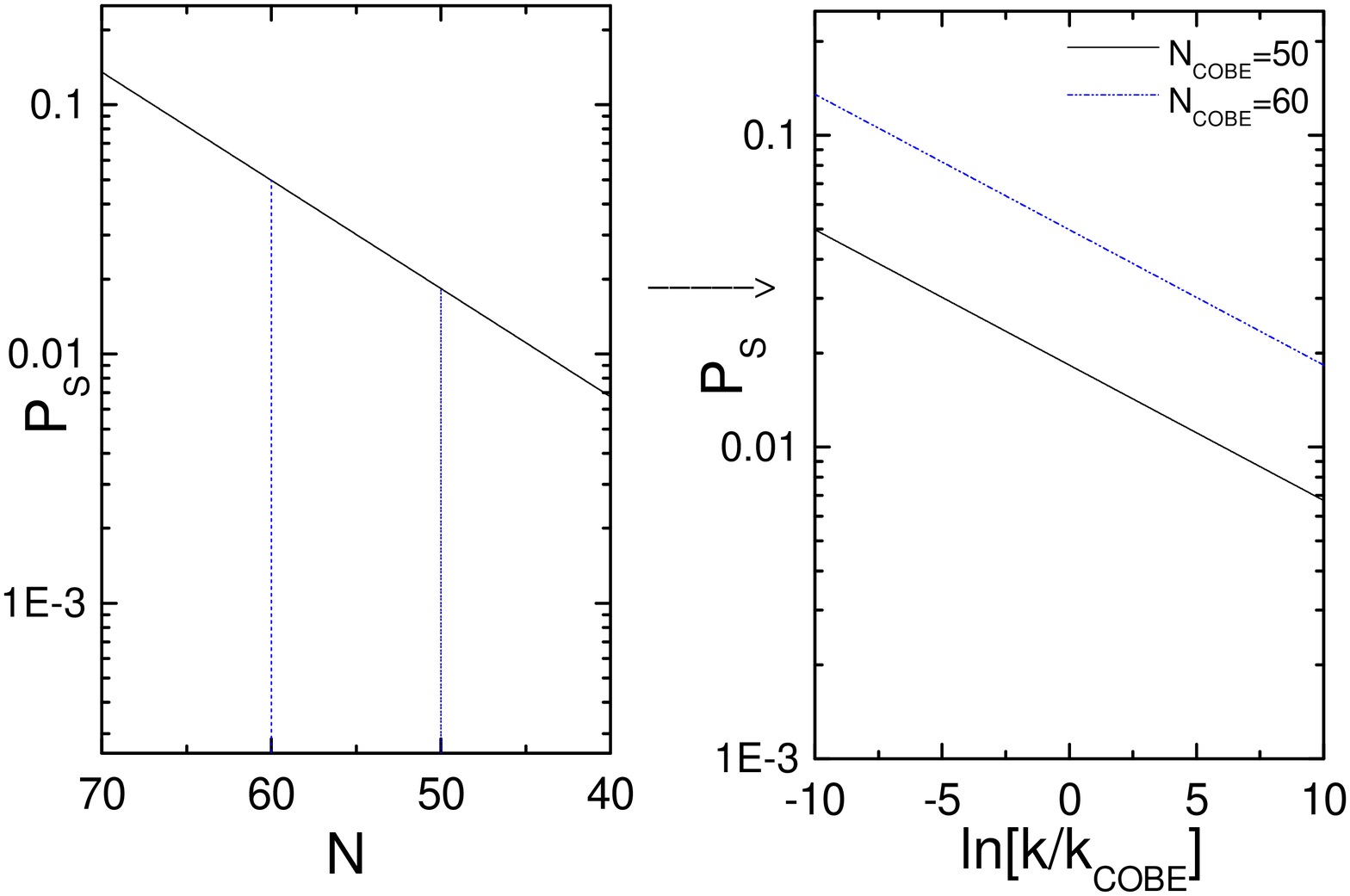}
\caption{Left: the primordial spectrum $P_S$ as a function of the
e-fold number $N$; Right: $P_S$ as a function of the wavenumber
$k$. The figure illustrates the effect on $P_S(k)$ when choosing
different $N_{COBE}$.}
\end{figure}

\begin{figure}
\includegraphics[scale=0.35]{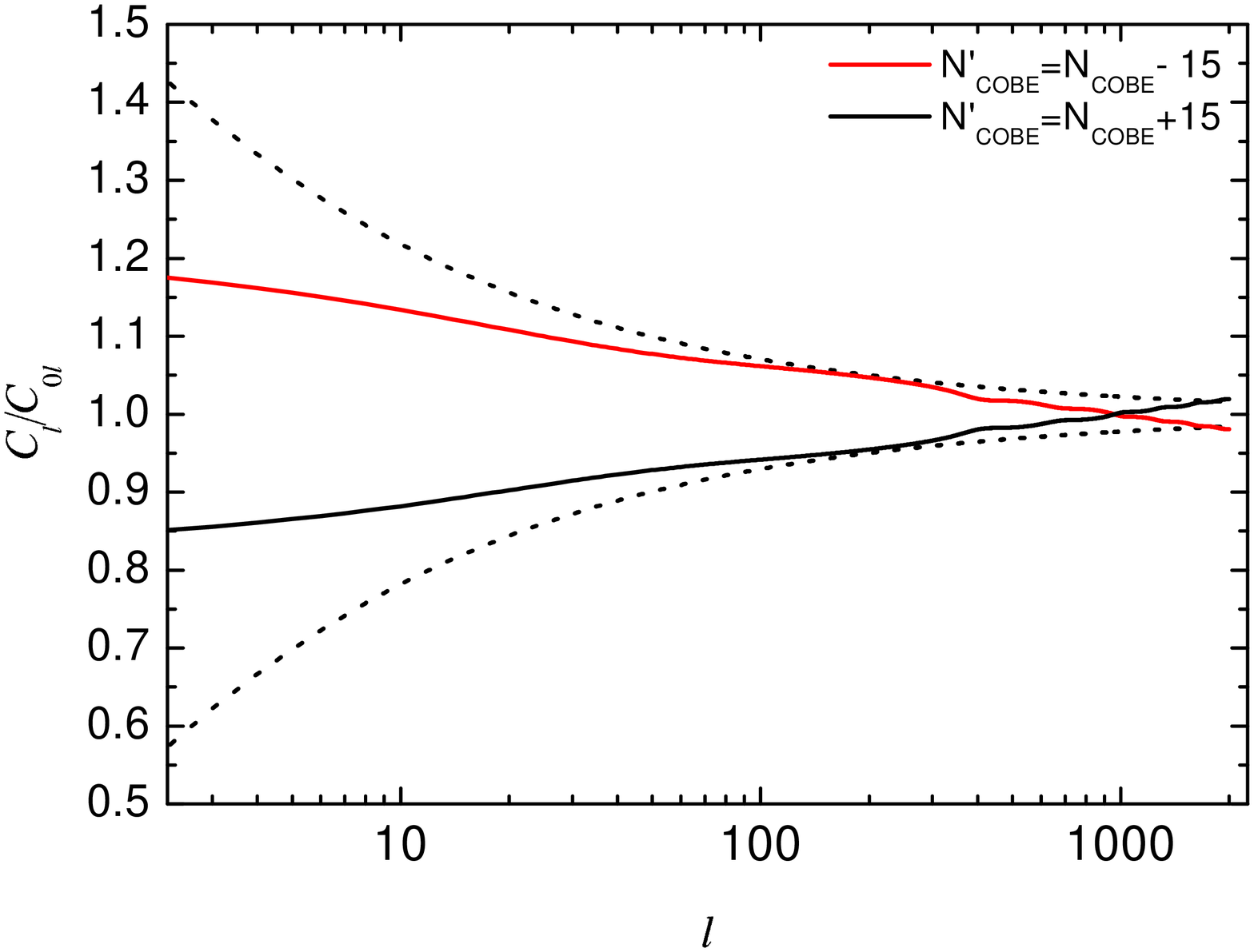}
\caption{The ratio of different CMB angular power spectra $C_l$
relative to $C_{0l}$. $C_{0l}$ stands for $n_S(k_{COBE})=0.91$ and
$dn_S/dlnk=-0.002$, the region between the two dashed lines are
allowed by its cosmic variance.}
\end{figure}

\begin{figure}
\includegraphics[scale=0.35]{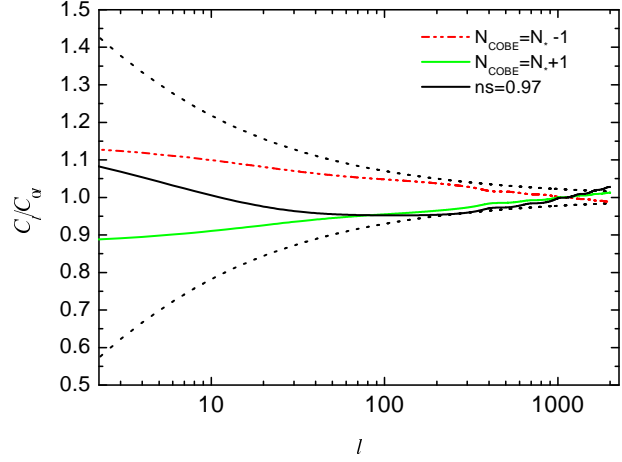}
\caption{$C_{0l}$ stands for $n_S(k_{COBE})=1.014$ and
$dn_S/dlnk=-0.022$.}
\end{figure}

\begin{figure}
\includegraphics[scale=0.5]{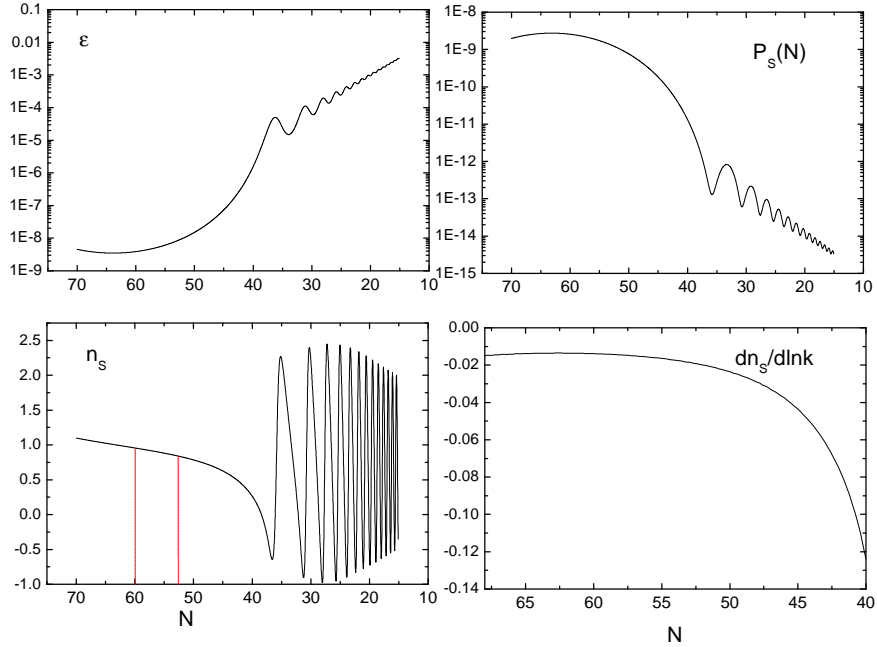}
\caption{The SR parameter $\epsilon$ , spectrum $P_S$, scalar
spectra index $n_S$ and its running $dn_S/dlnk$ with $N=300$,
$\delta= 5 \times 10^{-5}$ and $f=0.4 M_{Pl}$ for
$V(\phi)=V_0(1+\cos \frac{\phi}{f}+\delta \cos \frac{N\phi}{f})$,
the region within the two dashed lines is roughly allowed by the
model and current observations for $k_0=H_0$.}
\end{figure}

\begin{figure}
\includegraphics[scale=0.4]{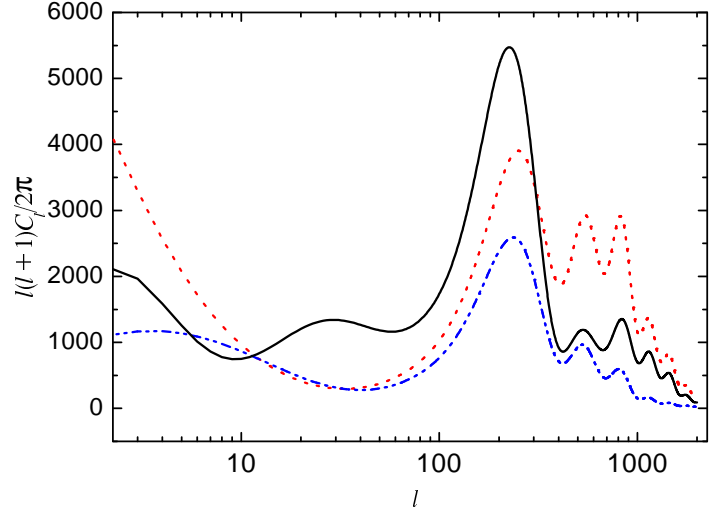}
\caption{CMB angular power spectra for choosing $N(k_{COBE})$
differently, assuming the existence of thermal inflation some time
after the inflation in Eq.11. From left top to
bottom, the lines stand for $N(k_{COBE})=37$ , 27 and 32
respectively.}
\end{figure}

\begin{figure}
\includegraphics[scale=0.4]{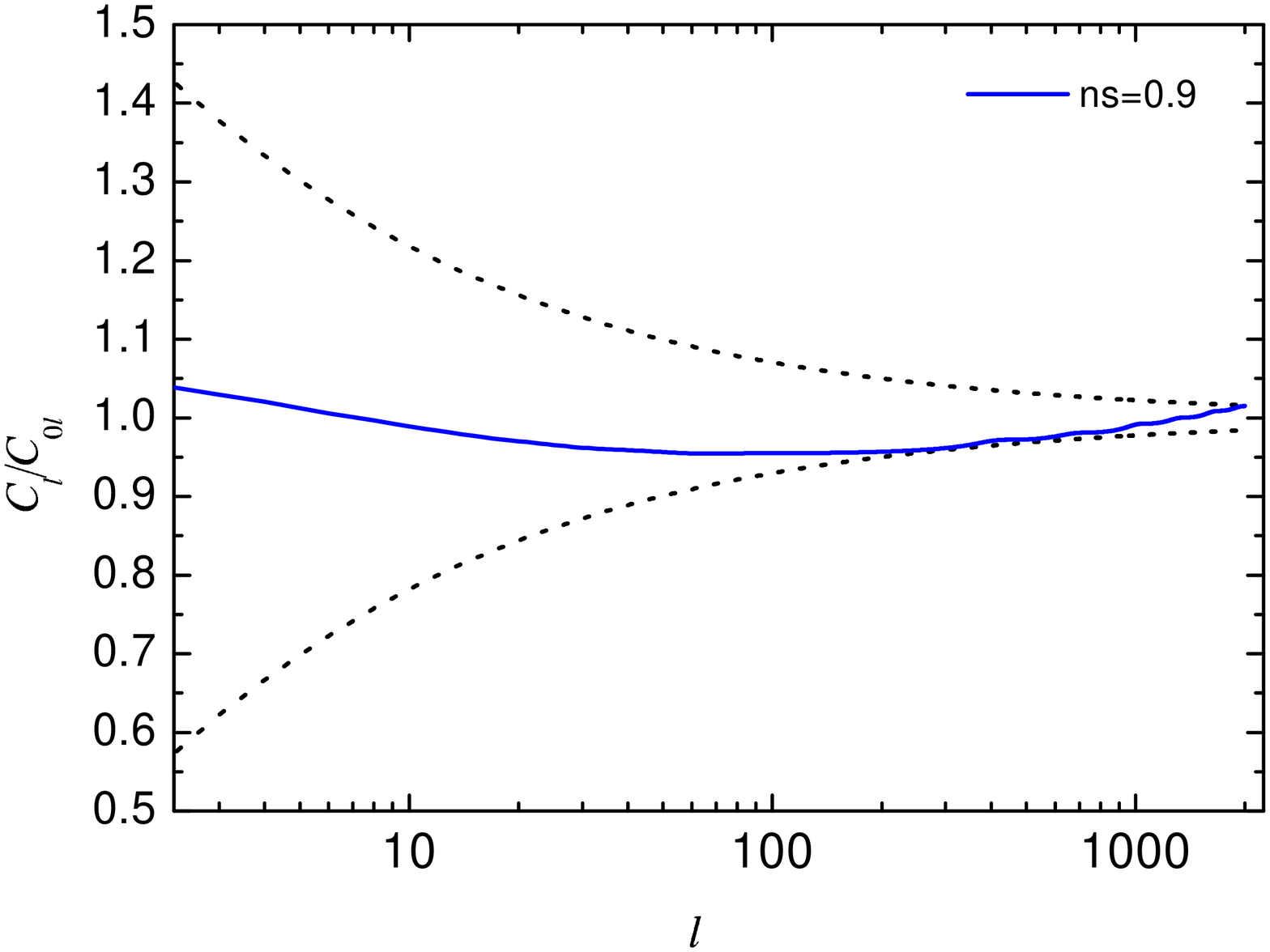}
\caption{$C_{0l}$ stands for $N(k_0)\approx 60$ for the $P_S(N)$
shown in Fig.4 .}
\end{figure}

\begin{thebibliography}{}


 \bibitem{Ma} C. Ma, E.Bertschinger , \APJ{455}{7}{1995}.


\bibitem{Lyth reports} D.H.Lyth and A.Riotto,
\PRPT{314}{1}{1999}.



\bibitem{Xiaom Wang}X. Wang, M. Tegmark, M. Zaldarriaga,
\PRD{65}{123001}{2002}.

\bibitem{L.Covi}D.H. Lyth and L.Covi, \PRD{62}{103504}{2000}.

\bibitem{Wangxiulian} X.Wang,B.Feng,M.Li, astro-ph/0209242.

\bibitem{Covi02} L.Covi, D. H. Lyth, A. Melchiorri,
hep-ph/0210395.

\bibitem{Bunn Liddle and White} E.Bunn , A.Liddle and
M.White ,\PRD{54}{5917}{1996}.


\bibitem{Thermal inflation}D. H. Lyth , E. D. Stewart, \PRD{53}{1784}{1996}.


\bibitem{Bunn and White} E.Bunn , M.White, \APJ{480}{6}{1997}.

\bibitem{white93}Martin White,Lawrence M. Krauss, Joseph Silk,\APJ{418}{535}{1993}.

\bibitem{ZS97}M.Zaldarriaga, U.Seljak, \PRD{55}{1830}{1997}.

\bibitem{KAA97}M.Kamionkowski, A.Kosowsky, A.Stebbins,
\PRD{55}{7368}{1997}.

\bibitem{TZ02}M. Tegmark, M. Zaldarriaga, astro-ph/0207047.







\bibitem{GSZ00} L. M. Griffiths, J. Silk, S. Zaroubi,
astro-ph/0010571.



\bibitem{LPS98} J. Lesgourgues, D. Polarski, A. A. Starobinsky,
\MNRAS{297}{769}{1998};\\
F. Atrio-Barandela, J. Einasto, V. M$\ddot{u}$ller, J. P. M$\ddot{u}$cket and A.
A. Starobinsky, \APJ{559}{1}{2001}.

\bibitem{KL00} M. Kamionkowski, A. R. Liddle,
\PRL{84}{4525}{2000};\\
 A. R. Zentner, J.S. Bullock
astro-ph/0205216 and reference therein.

\bibitem{Starobinsky92} A. A. Starobinsky, \JETPL{55}{489}{1992}.

\bibitem{Sarkar} S.Sarkar, \RPP{59}{1493}{1996}.

\bibitem{TM TT}See for an example, T.Moroi, T.Takahashi, \PLB{503}{376}{2001}.


\bibitem{Chung00} D. J. H. Chung, E. W. Kolb, A. Riotto and I. I. Tkachev, \PRD{62}{043508}{2000}.




\bibitem{BCLH02} C.P. Burgess, J.M. Cline, F. Lemieux, R. Holman,
 hep-th/0210233.



\bibitem{Seljak}U.Seljak, M.Zaldarriaga, \APJ{469}{437}{1996};\\
http://physics.nyu.edu/matiasz/CMBFAST/cmbfast.html.
\end{thebibliography}
\end{document}